\documentclass[aps,pra,twocolumn,showpacs]{revtex4}
\usepackage[pdfview=FitV,pdfstartview=FitV]{hyperref}
\usepackage{amsmath,amssymb,amsfonts,empheq,bm}
\usepackage{graphicx}
\usepackage{epsfig}
\usepackage{color}
\newcommand{\mbf}[1]{{\textbf #1}}

\begin{document}
\title{Roton confinement in trapped dipolar Bose-Einstein condensates}
\author{M. Jona-Lasinio, K. {\L}akomy and L. Santos}
\affiliation{Institut f\"ur Theoretische Physik, Leibniz Universit\"at, 30167 Hannover, Germany}
\date{\today}

\begin{abstract}
Roton excitations constitute a key feature of dipolar gases, connecting these
gases with superfluid helium. We show that the density dependence of the roton
minimum results in a spatial roton confinement, particularly relevant
in pancake dipolar condensates with large aspect ratios. We show that roton confinement plays
a crucial role in the dynamics after roton instability, and that arresting the instability may create a trapped roton gas revealed by
confined density modulations. We discuss the local susceptibility against
density perturbations, which we illustrate for the case of vortices. Roton
confinement is expected to play a key role in experiments.
\end{abstract}

\pacs{03.75.Kk 05.30.Jp and 67.85.-d}
\maketitle


\section{Introduction}
\label{sec:Introduction}

Dipolar gases have attracted growing attention in recent years. Quantum degenerate
gases of magnetic atoms as chromium~\cite{Griesmaier2005},
dysprosium~\cite{Lu2011}, and erbium~\cite{Aikawa2012} have already been
realized. Moreover, the preparation of heteronuclear molecules in their
ro-vibrational ground state~\cite{Ni2008} opens the path for the
creation of a degenerate gas of polar molecules, a goal currently pursued
by various groups worldwide~\cite{Wu2012}. Rydberg atoms provide yet another
possible realization of a highly polar gas~\cite{Gallagher2008}.

The rich physics of dipolar gases arises from
dipole-dipole interactions~(DDIs)~\cite{Baranov2008}. Dipolar Bose-Einstein
condensates~(dBECs) feature a geometry-dependent stability~\cite{Muller2011} and
a peculiar dispersion of the elementary excitations. Nonpolar BECs present the
usual Bogoliubov spectrum, with a linear~(phonon) dispersion at low
momenta and a quadratic dispersion at large
momenta~\cite{Stringari-Book}. Under proper conditions, dBECs
present a dispersion minimum at intermediate
momenta~\cite{Santos2003,Ronen2007} resembling the roton minimum of superfluid
He~\cite{Landau1947}.

Roton excitations are crucial in He, reducing the critical
superfluid velocity~\cite{Landau1941} and leading to density
modulations at defects
~\cite{Regge1972,Dalfovo1992,Pomeau1993,Berloff1999,Villerot2012}.
Remarkably, similar effects have also been predicted in dBECs~\cite{Wilson2010,
Yi2006,Wilson2008,Lu2010}. Moreover, the roton minimum is crucial for the
stability of a dBEC. When the minimum reaches zero energy, the dBEC becomes
unstable against finite-momentum excitations~(roton
instability)~\cite{Santos2003,Ronen2007,Wilson2009}, which differs fundamentally
from the usual phonon instability. 

\begin{figure}[t]
\centering
\includegraphics[clip=true,width=\columnwidth]{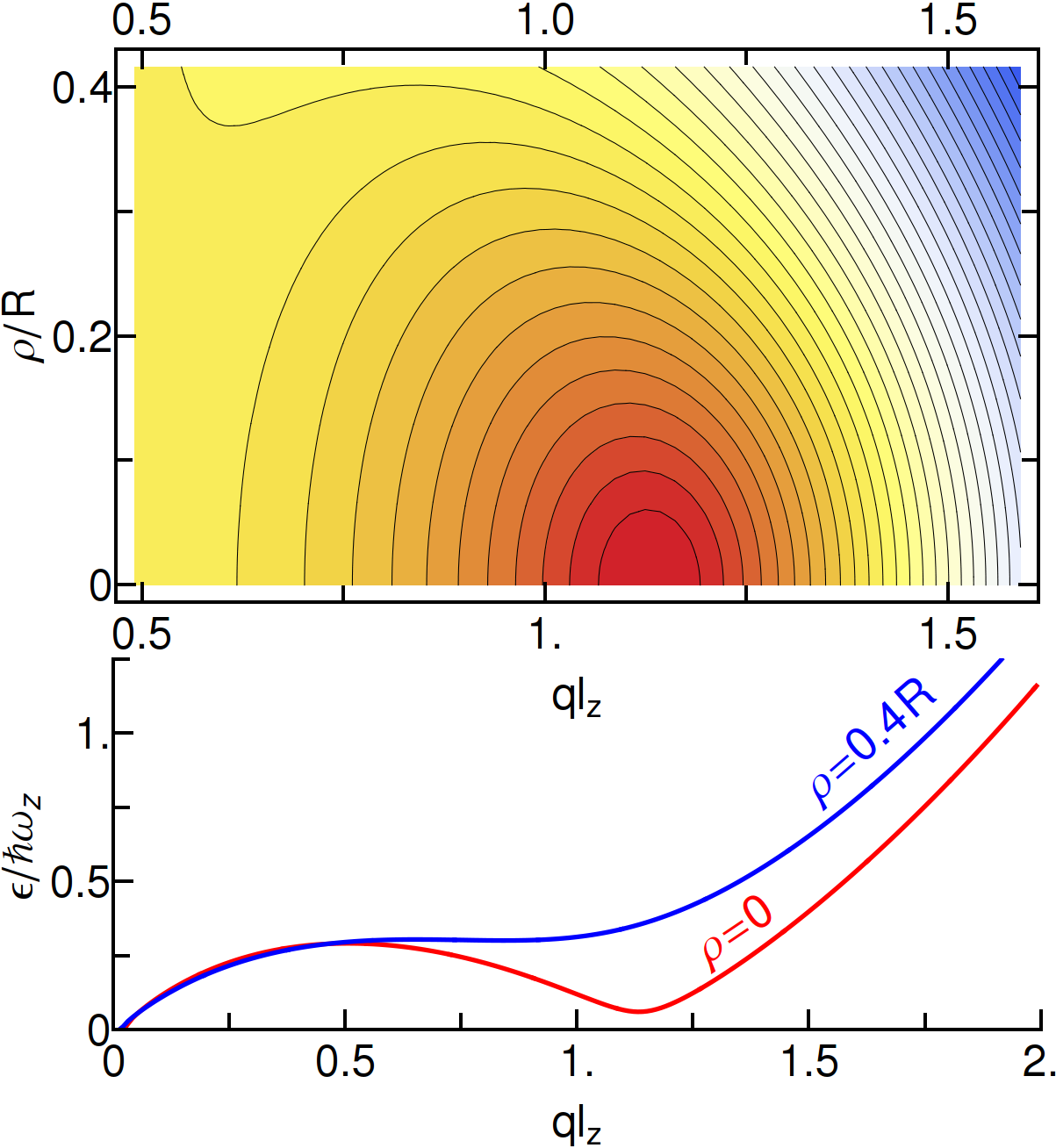}
\vspace*{-0.2cm}
\caption{(Color online)
(Top) Local spectrum of a BEC of $2\times 10^5$ Er atoms $\omega_z=2\pi\times 1$
kHz and $\lambda=40$.  Note a minimum~(dark red region in the
top panel) in both space and momentum. (Bottom) Change of the roton minimum for
two different radial positions. 
}
\label{fig:1}
\end{figure}

Roton properties in He may be controlled by means of
pressure~\cite{Dietrich1972}. Similarly, the roton minimum in dBECs depends on
interactions and therefore on density. In this paper we demonstrate that this
density dependence leads to a spatial roton confinement in trapped dBECs. Roton
confinement has been hinted in recent numerical
calculations~\cite{Wilson2010,Blakie2012} and resembles that of rotons at vortex
lines in He~\cite{Iguchi1972}.
We show that roton confinement is crucial 
to understand the roton instability in pancake traps.
We also discuss how arresting the instability may create a
trapped roton gas. We finally analyze other consequences of the roton
confinement, such as local susceptibility against density perturbations.

The paper is organized as follows. In Sec.~\ref{sec:Model} we present the system
under consideration.  In Sec.~\ref{sec:Homogeneous} we briefly review the main
results of Ref.~\cite{Santos2003}. The  key idea of local spectrum is introduced
in Sec.~\ref{sec:LocalSpectrum}. In  Sec.~\ref{sec:RotonWavefunction} we obtain
the wave functions of the confined rotonlike excitations. 
Section~\ref{sec:RotonInstability} is devoted to the key role played by roton
confinement in the dynamics following  roton instability, whereas
Sec.~\ref{sec:TOF} analyzes the main features of  the localized roton
instability in time-of-flight pictures. In Sec.~\ref{sec:ConfinedRotonGas} we
comment on the realization of a confined roton gas by means of  a temporal
destabilization of the condensate. In Sec.~\ref{sec:LocalSusceptibility} we
discuss the local susceptibility  associated with the idea of local spectrum,
focusing on the particular case of vortex lattices. Finally we summarize our
conclusions in Sec.~\ref{sec:Conclusions}.


\section{Model}
\label{sec:Model}

We consider a dBEC of $N$ bosons of mass $m$ and (electric or magnetic)
dipole moment $d$ oriented along $z$. The dBEC is in a pancake harmonic trap
$V_t(\mbf{r})$ of frequencies $\omega$ in the $xy$ plane and $\omega_z=\lambda
\omega$ along $z$, with $\lambda\gg 1$. The dBEC wave function $\phi(\mbf{r},t)$
obeys the nonlocal Gross-Pitaevskii equation~(GPE)~\cite{Baranov2008},
\begin{align}
i\hbar\frac{\partial}{\partial t}\phi(\mbf{r},t)=&
\left [ 
-\frac{\hbar^2\nabla^2}{2m}+ V_t(\mbf{r}) + g|\phi(\mbf{r},t)|^2 
\right ] \phi(\mbf{r},t) \nonumber \\
&+ \int d^3 r' V_{dd}(\mbf{r}-\mbf{r'}) |\phi(\mbf{r'},t)|^2 \phi(\mbf{r},t), 
\label{eq:GP}
\end{align}
where $g=4\pi\hbar^2 aN/m$ characterizes the short-range interactions, $a$ is
the $s$-wave scattering length,
$V_{dd}(\mbf{r})=\frac{Nd^2}{r^3}(1-3\cos^2\theta)$ is the DDI potential,
$\theta$ is the angle between $\mbf{r}$ and the $z$ axis, and $\int d^3 r
|\phi(\mbf{r},t)|^2=1$.


\section{Homogeneous case}
\label{sec:Homogeneous}

We first consider $\omega=0$, briefly summarizing the results of
Ref.~\cite{Santos2003}~(for more details, see the Appendix~\ref{App:A}). The
ground-state wavefunction is $\phi_0(z)\exp{(-i\mu t/\hbar)}$, where $\mu$ is
the chemical potential and $\phi_0(z)$ fulfills a one-dimensional (1D) local GPE
with a regularized coupling constant $g+g_d$, with $g_d=8\pi N d^2/3$. Assuming
a transverse Thomas-Fermi~(TF) profile $\phi_0(z)=\sqrt{n_0 \left ( 1-z^2/L^2
\right )}$, with $n_0$ the peak density, one obtains $\mu=(g+g_d) n_0$.
Excitations of energy $\epsilon$ are evaluated by substituting
$\phi({\boldsymbol{\rho}},z,t)=e^{-i\mu t/\hbar}\left [
\phi_0(z)+u(\boldsymbol{\rho},z)e^{-i\epsilon t/\hbar}-v(\boldsymbol{\rho},z)^*
e^{i\epsilon t/\hbar} \right ]$ ~[with $\boldsymbol{\rho}\equiv (x,y)$] into
Eq.~\eqref{eq:GP}, and keeping only linear terms in $u$ and $v$. The excitations
have a defined in-plane momentum ${\bf q}$, and hence $f_{\pm}
(\boldsymbol{\rho},z)\equiv \left [ u(\boldsymbol{\rho},z) \pm
v(\boldsymbol{\rho},z) \right ]=f_\pm (z) e^{i{\bf q}\cdot \boldsymbol{\rho}}$.
The lowest eigenenergy obtained from the Bogoliubov de Gennes~(BdG) equations
for each ${\bf q}$ builds the dispersion $\epsilon(q)$. Interestingly,
$\epsilon(q)$ may present a rotonlike minimum at intermediate $q$. Assuming
$qL\gg 1$ and $\mu E(q) |\beta -2|/(\beta+1)\lesssim \hbar^2\omega_z^2$~(with
$\beta\equiv g_d/g$ and $E(q)\equiv\hbar^2q^2/2m$) we obtain an approximate
expression of the dispersion,
$\epsilon_{h}^2(q,\mu)=E(q)^2-G(\beta)E(q)\mu+\hbar^2\omega_z^2$, with $G(\beta)
\equiv \frac{(\beta-2)(5\beta+2)}{3(1+\beta)(2\beta+1)}$, as well as the
associated eigenstate,  $f_+(z)\simeq \phi_0(z)$. The expression $\epsilon_h(q,
\mu)$ agrees well with the numerical $\epsilon(q)$ and shows that the roton
depth depends explicitly on $\mu$ and hence on density.


\section{Local spectrum}
\label{sec:LocalSpectrum}

Interesting insights about the roton physics for $\omega> 0$ are provided by the
concept of {\em local spectrum} $\epsilon(q, \rho)$, which we introduce here. We
compute from Eq.~\eqref{eq:GP} the ground-state $n_{0}(\mbf{r})=
|\phi_{0}(\mbf{r})|^2$, which in the TF regime is approximated by
$n_0(\mbf{r})={\tilde n}_0 \left ( 1-\rho^2/R^2-z^2/L^2 \right )$~\cite{footnote-biconcave}. We obtain for
each $\rho$ the $z$ profile $n_{0}^{1\text{D}}(z)=n_0(\mbf{r})/\int\! d z\:
n_0(\mbf{r})$ and the {\em local} chemical potential $\mu_{l}(\rho)$. Solving
the corresponding 1D BdG equations~\cite{Santos2003} we obtain
$\epsilon_{h}(q,\mu_{l}(\rho))$, which approximates the local spectrum
$\epsilon(q,\rho)$ consistently with the LDA. The local chemical potential
$\mu_l(\rho)$ decreases quadratically with $\rho$, and hence at the trap center
($\rho=0$) the roton energy is lowest: $\epsilon_r/\hbar\omega_z \equiv
\epsilon(q=q_r,\rho=0)/\hbar\omega_z
\simeq\sqrt{1-(G(\beta)\mu_l(0)/2\hbar\omega_z)^2}$.  The local spectrum presents
a minimum both in momentum, at $q_r l_z\simeq\sqrt{ G(\beta)\mu_l(0)/
\hbar\omega_z }$  (note that $q_r  l_z\sim 1$, with $l_z\equiv\sqrt{\hbar/m\omega_z}$,
see~\cite{Santos2003}), and in space, at $\rho=0$~(Fig.~\ref{fig:1}). For larger
$\rho$ the minimum becomes shallower and eventually disappears.  Hence,
remarkably, the inhomogenous BEC density results in a spatial {\em roton
confinement}. Around the minimum,
\begin{align}
\epsilon(q,\rho)\simeq\epsilon_r + \frac{\hbar^2(q-q_{r})^2}{2m_*}+\frac{1}{2}m_*\omega_*^2 \rho^2,
\label{eq:star}
\end{align}
where the effective roton mass  $m_*\equiv m \sqrt{1/\left (q_{r}l_z\right )
^{4}-1/4}$ and the effective harmonic frequency $\omega_*\equiv \omega_z \frac{mq_{r}
l_z^2}{m_*R\sqrt{2}}$ define the roton localization length
$l_*\equiv\sqrt{\hbar/m_*\omega_*}=2^{1/4}(R/q_{r})^{1/2}$. Note that $l_*/R\sim
\sqrt{l_z/R}$, and hence $l_*\ll R$ for $\lambda\gg 1$. Moreover, $q_{r} l_*
\sim \sqrt{R/l_z}\gg 1$ if $\lambda\gg 1$, which justifies the use of LDA
above. 

\begin{figure*}[t]
\centering
\includegraphics[clip=true,width=2.0\columnwidth]{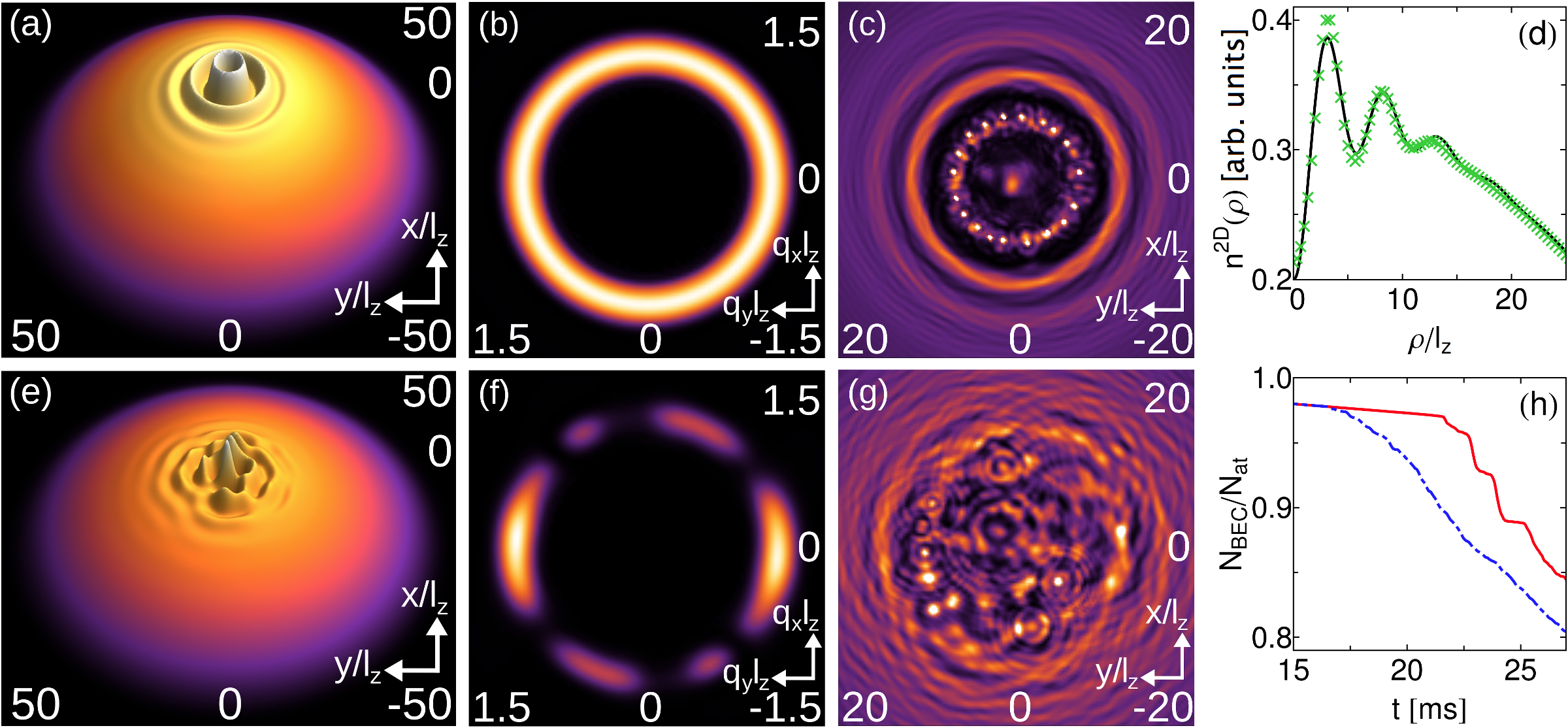}
\vspace*{-0.2cm}
\caption{
(Color online) Roton instability for $10^5$ Er atoms, with $\omega_z = 2\pi
\times 450$ Hz~($l_z\simeq 0.3$ $\mu$m), $\lambda=30$,
$a_i=8.49a_0>a_c=8.48a_0$ and $a_f=0$~(see text). (a) Column density
$n^{2D}(\boldsymbol{\rho})=\int n({\bf r}) dz$ showing concentric rings
($s=0$) formed after $t=19$ ms for small initial fluctuations~($\xi=10^{-10}$;
see text). (e) Modulational instability after $t=15.5$ ms consisting of several
$s$ states for large initial fluctuations~($\xi=10^{-6}$; see text). (b) and (f):
momentum distribution of (a) and (e) respectively (we have suppressed
the large peak at $q=\sqrt{q_x^2+q_y^2}=0$). (c) and (g):
post-collapse dynamics after $t=23$ ms for (a) and $t=19.5$ ms for (e),
respectively. (d) Radial cut of (a)~(green crosses) and theoretical
column density of the form $n^{2D}(\boldsymbol{\rho})\propto
(1-\rho^2/R^2)^{3/2} (1+A J_0(q_r \rho) e^{-\rho^2/2l_*^2})$~(solid black line),
with $R\simeq 51\,l_z$, $A\simeq -0.4$, $q_r l_z\simeq 1.25$, and $l_*=2^{1/4}
(R/q_r)^{1/2}\simeq 7.6\,l_z$. The value of $q_r$ is calculated using
the local spectrum picture. (h)
Remnant atoms for the cases (a) (solid red line) and (e) (dashed blue line).}
\label{fig:2}
\end{figure*}


\section{Localized roton wavefunction}
\label{sec:RotonWavefunction}

We now calculate the localized roton wave functions using the LDA and the
formalism of Ref.~\cite{Santos2003}~(for more details, see the
Appendix~\ref{App:A}). Assuming $l_*\ll R$, we approximate $f_+
(\boldsymbol{\rho},z)\simeq F(\boldsymbol{\rho})\phi_0(\boldsymbol{\rho},z)$,
where $F(\boldsymbol{\rho})$ has a narrow momentum distribution ${\tilde F}({\bf
q})$ centered around $q_r$ with a width $\delta q \propto 1/l_*\ll q_r$. The
finite width of ${\tilde F}({\bf q})$ must be now considered and as a result
${\tilde F}({\bf q})$ fulfills the eigenvalue equation $\epsilon^2 {\tilde
F}({\bf q}) \simeq [ \epsilon_r + \hat H ]^2 {\tilde F}({\bf q})$, with $\hat
H\equiv  \frac{\hbar^2}{2m_*}(q-q_{r})^2-\frac{1}{2}m_*\omega_*^2 \nabla_{\bf
q}^2$. Interestingly, $\hat H$ resembles the Hamiltonian of a trapped BEC in the
presence of spin-orbit coupling~\cite{Stanescu2008,Sinha2011}, where the
Rashba-like dispersion $\sim(q-q_{r})^2$ acts as a ringlike potential in ${\bf
q}$ space.  The eigenfunctions of $\hat H$, $\eta_{n,s}(q)
e^{is\varphi}/\sqrt{q}$, fulfill
\begin{equation}
\left [ \frac{E_{n,s}}{\hbar\omega_*} - \frac{s^2-\frac{1}{4}}{2(ql_*)^2} \right ] \eta_{n,s}= 
\left [ -\frac{1}{2l_*^2} \frac{d^2}{dq^2} +\frac{l_*^2}{2}(q-q_{r})^2 \right ]\eta_{n,s}.
\end{equation}
For $q_rl_*\gg 1$, we expand around $q\simeq q_r$, obtaining the
eigenenergies $E_{n,s}/\hbar\omega_*\simeq (s^2-1/4)/(2(q_rl_*)^2)+n+1/2$,
characterized by the angular momentum $s$ around the Rashba-like ring, and
the radial harmonic excitations, $n$, with frequency $\omega_*$. The lowest
roton states have $n=0$, being in real space of the form 
$\psi_s(\boldsymbol{\rho})\sim e^{i s \varphi}e^{-\rho^2/2l_*^2}J_s(q_r\rho)$,
with $J_s$ the Bessel function.


\section{Local roton instability}
\label{sec:RotonInstability}

\begin{figure*}[t]
\centering
\includegraphics[clip=true,width=2\columnwidth]{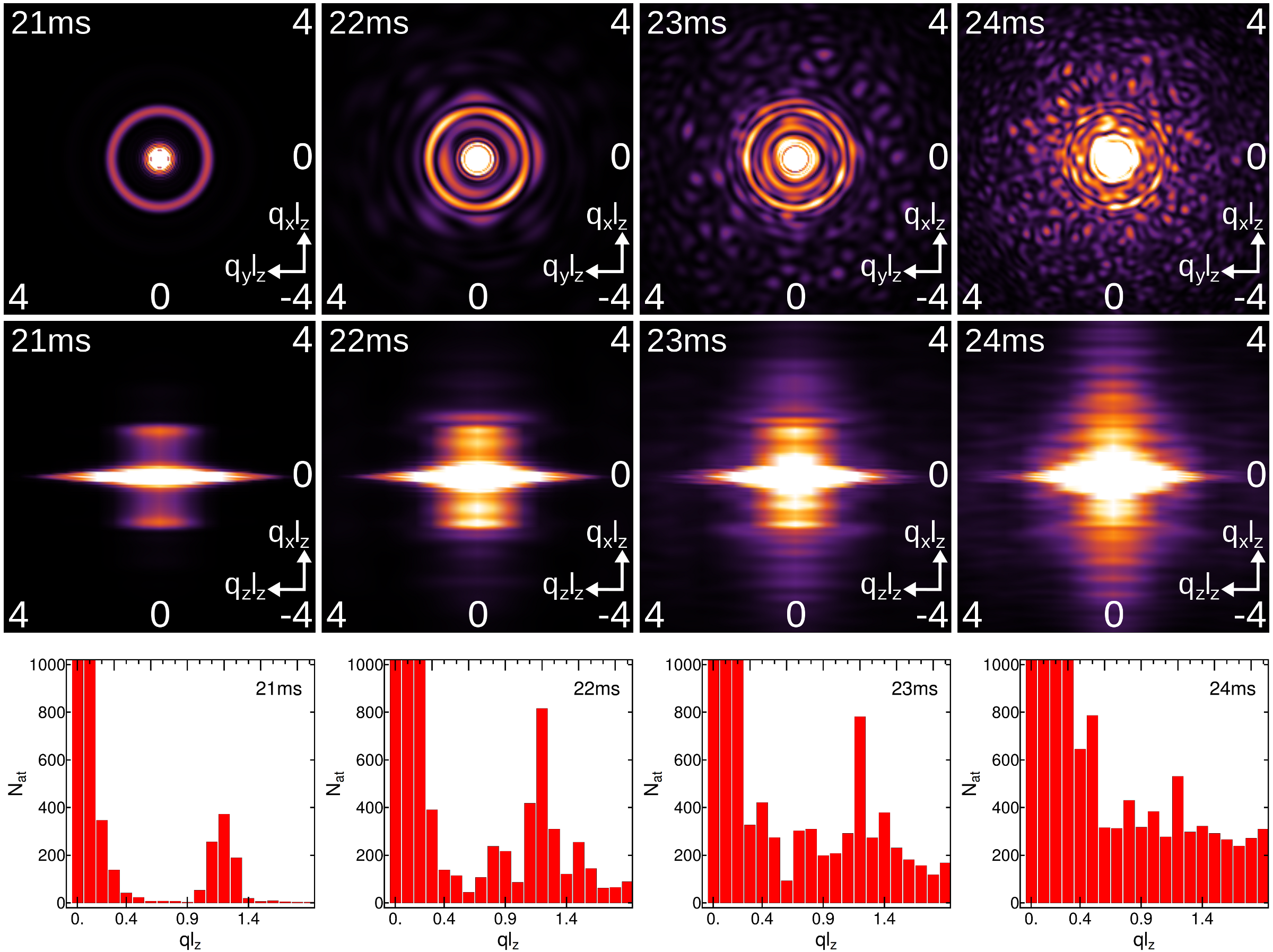}
\vspace*{-0.2cm}
\caption{
(Color online) Momentum distribution for different times for the same case
depicted in Fig.~\ref{fig:2}(a), as it would be revealed by a TOF
experiment. In the top panel we show the $q_x, q_ y$ momentum distribution
(integrated over $q_z$) and in the middle panel the $q_x, q_z$ momentum
distribution (integrated over $q_y$). The bottom panel shows the integrated
population at different intervals of the radial momentum
$q=\sqrt{q_x^2+q_y^2}$.}
\label{fig:5}
\end{figure*}

The localized states $\psi_s$ are crucial for the dynamics following roton
instability~\cite{footnote0}.  We consider a stable BEC prepared with an initial scattering length
$a_{i}> a_c$. The critical $a_c$ for the onset of 
instability depends nontrivially on $\lambda$ and $g_d$~\cite{Baranov2008}; for
the parameters of Fig.~\ref{fig:2} it is $a_c=8.48\,a_0$, with $a_0$ the Bohr
radius. We are interested in the instability dynamics after a sudden quench to
$a_f<a_c$. Unstable modes lead to a modulational instability characterized by a
growing density perturbation $\delta n ({\bf r,t})\equiv n({\bf r},t)-n_0({\bf
r}) \propto \sqrt{n_0({\bf r})} \Re(f_-({\bf r}))$. As for the
$\omega=0$ case, $f_- \simeq (E(q_r)/\epsilon) f_+ \propto \sqrt{n_0({\bf r})}
F(\boldsymbol{\rho})$. Hence the most unstable roton modes result in a localized
modulation at the trap center, $\delta n ({\bf r})/n_0({\bf r}) \propto
\Re(\psi_s(\boldsymbol{\rho}))$. 

The most unstable mode is $\psi_0(\boldsymbol{\rho})$, but other modes may
contribute to the instability due to the small energy difference
between low-$s$ levels [$\sim \hbar \omega_*/(q_rl_*)^2$]. As a nontrivial
consequence of that, the density pattern that develops after the quench is
influenced by the initial population of the
excitations exponentially amplified during the destabilization. We
mimic this dependence by considering a small initial seeding
$\phi({\bf r},t=0)=\phi_0({\bf r})e^{i\chi({\bf r})}$, where $\phi_0$ is the
ground state calculated for $a_i$, and $|\chi({\bf r})|/\pi<\xi$ is a random
phase sampled from a homogeneous uniform distribution with a variable amplitude
$\xi$. Although this allows us to discuss the possible collapse scenarios, the
actual amplitude of the initial fluctuations depends on $a_i$ and on the
temperature $T$, and its analysis lies beyond the scope of this
paper~\cite{footnote-T}. 

Figures~\ref{fig:2} show the results of our simulations of Eq.~\eqref{eq:GP} for
an erbium dBEC~\cite{footnote-otherspecies}. For a small initial population of
the unstable modes, the modulation instability proceeds at a sufficiently slow
pace and $\psi_0(\boldsymbol{\rho})$ dominates. As a result, a localized pattern
of concentric rings develops~[Fig.~\ref{fig:2}(a)], in excellent agreement with
$\delta n ({\bf r})/n_0({\bf r}) \propto \psi_0(\boldsymbol{\rho})$; see
Fig.~\ref{fig:2}(d)~\cite{footnote-rings}. The corresponding momentum distribution presents a
Rashba-like ring~[Fig.~\ref{fig:2}(b)]~\cite{footnote-k0}. In contrast, for larger
initial fluctuations the pattern growth is too fast to select $\psi_0$ only, and
the created density pattern results from a~(shot-to-shot dependent) linear
combination of different $\psi_s$, being characterized by a superposition of
eccentric collapse centers~[Fig.~\ref{fig:2}(e)]. The corresponding momentum
distribution still presents a ringlike structure but with an azimuthal
modulation arising from the linear combination of various
$\psi_s$~[Fig.~\ref{fig:2}(f)]. 

The global~(phononlike) collapse studied in chromium and erbium
dBECs~\cite{Lahaye2008,Aikawa2012} results in large atom losses and in a
$d$-wave pattern in time-of-flight~(TOF) pictures. Remarkably, the roton
instability discussed above leads to a very different collapse dynamics.
Three-body losses become crucial in the collapse dynamics and we included them
by adding
$-i\hbar\frac{L_3}{2}N^2|\phi(\mbf{r},t)|^4\phi(\mbf{r},t)$~\cite{Lahaye2008} to
Eq.~\eqref{eq:GP}, with a loss rate $L_3=10^{-28}$cm$^{-6}$s$^{-1}$.  The
concentric rings appearing in Fig.~\ref{fig:2}(a) eventually undergo a
sequential collapse and azimuthal instability, starting from the inner~(denser)
ones towards the outer ones~[Fig.~\ref{fig:2}(c)] leading to a step-like atom
number decrease~[Fig.~\ref{fig:2}(h)]. The superimposed eccentric collapse
centers  appearing in Fig.~\ref{fig:2}(e) lead to a complex post-collapse behavior
with characteristic mutually interfering jets expelled out of each local
collapse center~[Fig.~\ref{fig:2}(g)]. In this case the atom decrease is
smooth~[Fig.~\ref{fig:2}(h)].

\section{Time-of-flight pictures}
\label{sec:TOF}

Although the most straightforward way of studying roton confinement is of
course given by in situ measurements~\cite{footnote-Axel}, TOF pictures are
expected to show  a clear difference as well, when compared to the cloverleaf
pattern characteristic of the phonon collapse (as observed in
chromium~\cite{Lahaye2008} and, more recently, in erbium~\cite{Aikawa2012}). 
More specifically, TOF pictures may reveal clear traces of the Rashba-like
ring characterizing the localized rotonlike excitations.

Due to the fast expansion after release, TOF pictures are expected to
reproduce well the in situ momentum distribution at the time of releasing. 
We have calculated the momentum distribution of the condensate at different
stages during the instability dynamics. Our results for the same case studied
in Fig.~\ref{fig:2}(a) are depicted in Fig.~\ref{fig:5}, where we show
both the column density on the $q_x$-$q_y$ plane (i.e. integrated along
$q_z$) and that on the $q_x$-$q_z$ plane (i.e. integrated along $q_y$). We
show in particular the results obtained during the first stages of the
collapse dynamics. For the parameters of Fig.~\ref{fig:2}(a),  the
Rashba-like ring is clearly visible during the first stages of the collapse,
being destroyed later on due to the subsequent collapse of the spatial
ringlike density modulations.  As a result the ring TOF pattern is washed
out. In later stages TOF pictures are characterized by the appearance of
large momentum excitations on the $xy$ plane~(induced by the collapse of the
local rings) which becomes visible in the $q_x$-$q_z$ distribution as
pronounced jets along $x$.

The low-momentum peak always appears saturated in our pictures. It was
eliminated from Figs.~\ref{fig:2}(b) and~\ref{fig:2}(g), since we were only
interested in showing the Rashba-like feature in momentum space. The large
low-momentum peak is the result of two effects. The first one is purely
geometric. The $q=0$ peak is always  relatively strong, since it is not spread
in the angular variable $\varphi$ (as it is the ring feature discussed above).
The second reason is more physical, and it is linked with the  local nature of
the collapse. Since only a small fraction of the atoms is actually participating
in the modulational instability and in the subsequent collapse, a large fraction
of  atoms remains at low momentum. Note that the low momentum peak also broadens
due to the generation of low energy phonons during the collapse. As a result,
the  low-momentum peak dominates the TOF pictures. This is  in itself a clear
difference with respect to the phonon collapse, where the  initial low-momentum
peak transforms into a cloverleaf pattern due to the global nature of the
collapse.

However, although the relative number of particles produced in the
Rashba-like ring may be small, the actual absolute number of particles may be
certainly sizable.  In Fig.~\ref{fig:5} (bottom panel) we depict at various
times the number of particles at different radial momenta.  This plot shows
that a significant number of particles is found in the ring before the ring
feature is eventually washed out (up to $1500$ particles at $21.5$ ms). 

\begin{figure}[t]
\centering
\includegraphics[clip=true,width=\columnwidth]{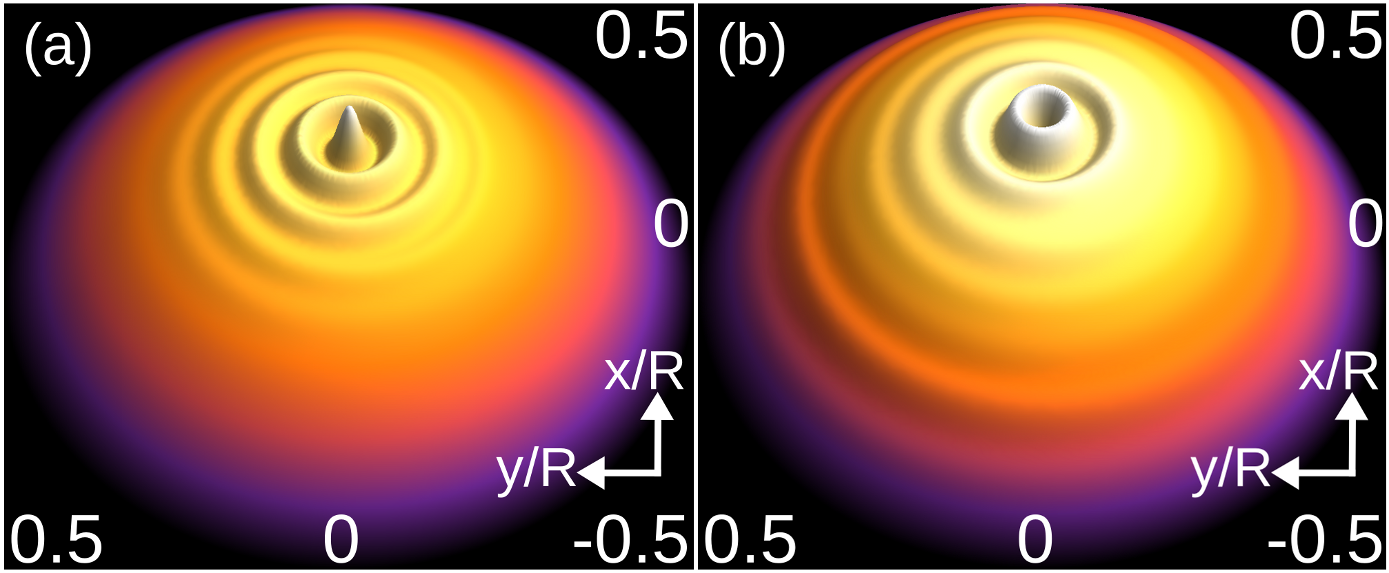}
\caption{
(Color online) Same parameters as in Fig.~\ref{fig:2}(a) but arresting the
collapse after $16$ ms with a quench of $a$ up to $8.55\, a_0 > a_c$. Snapshots
of the central region right after the quench (a) and $80$ ms later (b). Note a
clear density modulation confined at the trap center.
}
\label{fig:3}
\end{figure}

\begin{figure}[t]
\centering
\includegraphics[clip=true,width=\columnwidth]{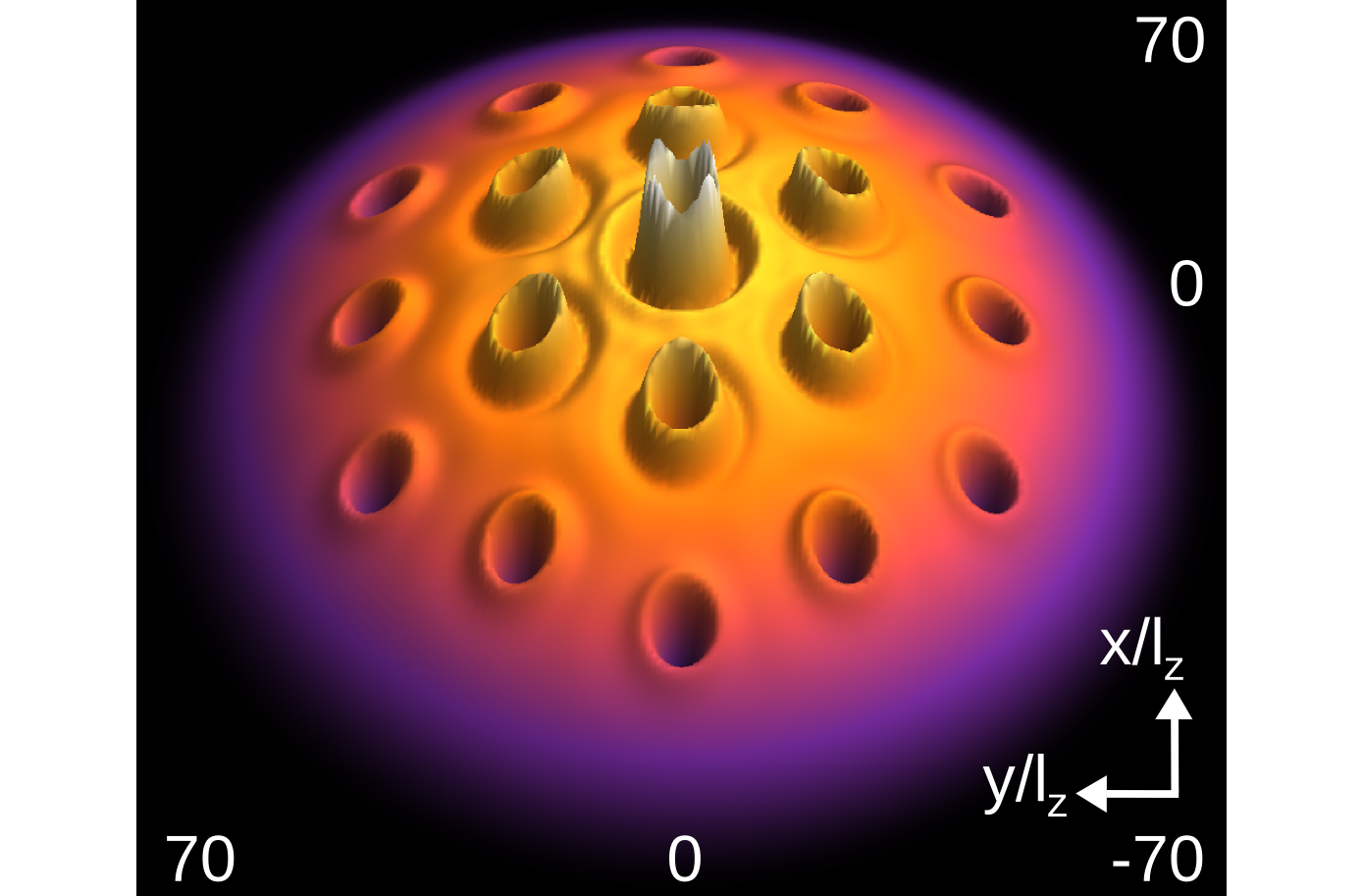}
\caption{
(Color online) Vortex lattice for a BEC of $N=10^5$ Er atoms at the threshold of
the roton instability. Parameters as in Fig.~\ref{fig:1} with a rotational 
frequency $0.3\,\omega$. 
}
\label{fig:4}
\end{figure}

\section{Confined roton gas}
\label{sec:ConfinedRotonGas}

Interestingly, the roton instability may be used to create a confined roton gas.
The density pattern obtained after quenching down to $a_f<a_c$ may be
interpreted as the growth of the roton population. At the initial stage of the
instability, the roton population is small compared to $N$ and we may neglect
condensate depletion or roton-roton interactions. Once the roton gas is
populated, we stabilize it by quenching up to $a>a_c$. Since
$m_*$ and $\omega_*$ do not vary significantly around the instability threshold,
the created rotons (density patterns) remain confined at the trap center after the
quench, as shown in Fig.~\ref{fig:3}.

\section{Local susceptibility}
\label{sec:LocalSusceptibility}

A deep roton minimum induces close to a perturbation a large
susceptibility against the formation of density modulations with the roton wavelength. 
This well-known effect in
He~\cite{Regge1972,Dalfovo1992,Pomeau1993,Berloff1999,Villerot2012} is also
relevant for dBECs~\cite{Wilson2008}. The local spectrum picture implies a
spatially dependent susceptibility, which we illustrate for the case of
vortices~\cite{footnote3}. Vortex cores present a craterlike shape in the case
of a deep roton minimum~\cite{Yi2006, Wilson2008} that disappears when the
minimum is shallow or absent. Therefore, vortices at different positions in a
trapped BEC present a different core profile. This is shown for a vortex lattice in Fig.~\ref{fig:4}, 
where we depict the ground state of an erbium BEC rotating
around the $z$ axis with an angular frequency $\Omega=0.3\,\omega$. Note that core
modulations at the trap center disappear for vortices at the boundary.

\section{Conclusions}
\label{sec:Conclusions}
In summary, we have shown that an inhomogeneous trapping in pancake dBECs
with large aspect ratios leads to a spatial roton confinement which is crucial
to understanding the roton instability. The roton dispersion has not yet been
observed experimentally, being currently a major goal pursued by several groups.
Roton confinement is expected to play a key role in these experiments, since
harmonic traps are typically employed and large aspect ratios are needed to
study the roton dispersion. In addition to the local susceptibility discussed
above~\cite{footnote3}, roton confinement should be carefully considered when
measuring the critical superfluid velocity, performing Bragg
scattering~\cite{footnote4}, or analyzing finite temperature physics, which may
be very interesting since the thermal roton cloud is expected to localize at the
center of the trap.

\acknowledgments We acknowledge funding by the German-Israeli Foundation, and
the DFG~(Grant No. SA1031/6 and Excellenzcluster QUEST). 

\appendix

\section{Derivation of the localized roton wave functions}
\label{App:A}
\subsection{Homogeneous $xy$ case ($\omega=0$)}

We briefly summarize in this section the main formalism developed in
Ref.~\cite{Santos2003}. For $\omega=0$, the ground state is of the form
$\phi_0(z)e^{-i\mu t/\hbar}$, where  $\mu$ is the chemical potential and
\begin{equation}
\left \{ -\frac{\hbar^2}{2m}\nabla^2+\frac{m}{2}\omega_z^2z^2+(g+g_{d})
|\phi_0(z)|^2-\mu\right \}\phi_0(z)=0.
\label{GPE1D}
\end{equation}
We assume $(g+g_d)>0$. For $\mu\gg\hbar\omega_z$ the condensate presents a
TF density profile, $|\phi_0(z)|^2= n_0(1-z^2/L^2)$, with
$n_0=\mu/(g+g_d)$ being the maximum density, and $L=\sqrt{2\mu/m\omega_z^2}$
the TF radius. 

Excitations of energy $\epsilon$ are evaluated by substituting  
$\phi({\boldsymbol{\rho}},z,t)=e^{-i\mu t/\hbar}\left [
\phi_0(z)+u(\boldsymbol{\rho},z)e^{-i\epsilon t/\hbar}-v(\boldsymbol{\rho},z)^* 
e^{i\epsilon t/\hbar} \right ]$ ~[with $\boldsymbol{\rho}\equiv (x,y)$] into the
3D nonlocal GPE [Eq.~\eqref{eq:GP} of our Letter], and keeping only linear terms in the
Bogoliubov amplitudes $u$ and $v$. For $\omega=0$, the excitations  have a
defined in-plane momentum ${\bf q}$, and  we may write  $f_{\pm}
(\boldsymbol{\rho},z)\equiv \left [ u(\boldsymbol{\rho},z) \pm
v(\boldsymbol{\rho},z) \right ]=f_\pm (z) e^{i{\bf q}\cdot \boldsymbol{\rho}}$. 
The BdG equations read
\begin{align} 
\epsilon f_{-}&=\frac{\hbar^2}{2m} \left [
-\frac{d^2}{dz^2}+q^2+\frac{\nabla^2\phi_0}{\phi_0}\right ] f_{+}
\equiv H_{kin}f_+, 
\label{BdGa} \\
\epsilon f_{+}&=H_{kin}f_{-}+H_{int}[f_-],  \label{BdGb} 
\end{align}
with
\begin{align}
H_{int}[f_-]&=2(g_{d}+g)f_{-}(z)|\phi_0(z)|^2-(3/2)g_d\,q\,
\phi_0(z) \nonumber \\ 
&\times\int_{-\infty}^{\infty}
d z'\,f_{-}(z')\phi_0(z')\exp{(-q|z-z'|)}.  
\label{Hint}
\end{align}
For each $q$ we calculate the eigenenergies. The lowest one, $\epsilon(q)$
builds the dispersion law.

The most interesting behavior occurs for $qL\gg 1$. We may now introduce
$f_+(z)=W(z)\phi_0(z)$. Expressing $f_-$ through $W$ from Eq.~\eqref{BdGa},  we
substitute it into Eq.~\eqref{BdGb} and integrate over  $dz'$ in $H_{int}[f_-]$ 
as the main contribution to the integral comes from a narrow range of distances
$|z'-z|\sim 1/q$. This yields~(here $\chi=z/L$)    
\begin{widetext}
\begin{eqnarray}      
&&\hbar^2\omega^2\left[\frac{1}{2}(1-\chi^2)\frac{d^2}{d\chi^2}-\left(1+\frac{3}{2(1+\beta)}
\right)\chi\frac{d}{d\chi}\right]W(\chi)  \nonumber \\
&=&-\left[\epsilon^2-E(q)^2-\frac{2\beta-1}{1+\beta}\mu
E(q)(1-\chi^2)-\frac{3\hbar^2\omega^2}{2(1+\beta)}\right]W(\chi)=0,   
\label{eigenmode}
\end{eqnarray}   
\end{widetext}
where $E(q)=\hbar^2q^2/2m$. Here we omitted terms of the order of
$E(q)\hbar^2\omega^2/\mu$ and $\hbar^4\omega^4/\mu^2$, since they are small
compared to either $\hbar^2\omega^2$ or $E(q)^2$. For each mode of the confined
motion (along the $z$ direction), the solution of Eq.~\eqref{eigenmode} can be
written as an expansion series in Gegenbauer polynomials $C^{\lambda}_n(\chi)$,
where $\lambda=(4+\beta)/2(1+\beta)$, and $n\geq 0$ is an integer.  For $\mu
E(q)|2\beta-1|/(1+\beta)\alt\hbar^2\omega^2$ the lowest solution is of the form 
$W(\chi)\simeq 1+\sum_{n>0} a_n C_n^\lambda (\chi)$, with small amplitudes $a_n$
of Gegenbauer polynomials of higher order, while the lowest-eigenenergy (and hence the
dispersion) is given  by
\begin{equation}
\epsilon_h^2(q,\mu)=E(q)^2-G(\beta) E(q)\mu
+\hbar^2\omega_z^2,
\end{equation}
with $G(\beta)\equiv \frac{(\beta-2)(5\beta+2)}{3(1+\beta)(2\beta+1)}$.

\subsection{Trapped $xy$ case}
We consider in the following $\omega>0$, with $\lambda=\omega_z/\omega\gg 1$.
Due to the $xy$ confinement, we  cannot employ anymore a plane wave basis for
the Bogoliubov amplitudes, as we did above. However, we assume that (i) the 
Bogoliubov amplitudes are, in momentum space, strongly peaked around a momentum
$q\gg 1/R$ and (ii) the Bogoliubov amplitudes are localized at the center of the
trap, in a region with a radius much smaller than  $R$. These assumptions must be
checked self-consistently, but we anticipate that they are fulfilled for a
sufficiently large aspect ratio $\lambda\gg 1$. If this is the case, we may
proceed as  for the homogeneous case, but taking into account that (a) the
chemical potential is now $\rho$ dependent, and (b) the Bogoliubov amplitudes
are not plane waves anymore. Assuming the above conditions are met, we may
write  $f_+(\boldsymbol{\rho},z)\simeq F(\boldsymbol{\rho}) W(x) \phi_0(
\boldsymbol{\rho},z)$ (now $x^2=\rho^2/R^2+z^2/L^2$), where  the function
$F(\boldsymbol{\rho})$ has a narrow momentum distribution $F({\bf q})$ peaked at
$q_r$ with a momentum width $\delta q\propto 1/l_*  \ll q_r$ and $\phi_0(
\boldsymbol{\rho},z)$ is the ground state. Taking into account (a) and (b), we
rewrite Eq.~\eqref{eigenmode}  in the form
\begin{widetext}
\begin{eqnarray}      
&&\hbar^2\omega^2 
\left[\frac{1}{2}(1-x^2)\frac{d^2}{dx^2}-\left(1+\frac{3}{2(1+\beta)}\right)x\frac{d}{dx}\right]
W(x)F(\boldsymbol{\rho}) \nonumber \\
&=&-\left[\epsilon^2-E(\hat {\bf q})^2-\frac{2\beta-1}{1+\beta}\mu
E(\hat {\bf q})(1-x^2)-\frac{3\hbar^2\omega^2}{2(1+\beta)}\right]W(x) F(\boldsymbol{\rho})=0,   
\label{eigenmode-LDA}
\end{eqnarray}  
\end{widetext}
where $\hat {\bf q}\equiv -i \boldsymbol{\nabla}$. The lowest energy eigenstates
still fulfill $W(x)\simeq 1$, but now we must keep explicitly the spatial
dependence of $F(\boldsymbol{\rho})$,
\begin{equation}    
\epsilon^2 F(\boldsymbol{\rho})=\left [ E(\hat {\bf q})^2-G(\beta) E(\hat {\bf q})\mu(\rho)
+\hbar^2\omega_z^2 \right ] F(\boldsymbol{\rho}).
\end{equation}
Expanding around the roton minimum in the local spectrum, and moving to momentum
space ($\boldsymbol\rho=i\boldsymbol{\nabla}_q$) we obtain
\begin{equation}    
\epsilon^2 \tilde F({\bf q})=\left [ \epsilon_r + 
\frac{\hbar^2( q-q_{r})^2}{2m_*}-\frac{1}{2}m_*\omega_*^2 \nabla_q^2 \right ]^2
 \tilde F({\bf q}),
\end{equation}
where $\tilde F({\bf q})$ is the Fourier transform of $F(\boldsymbol{\rho})$.
Hence $\tilde F({\bf q})$ are the eigenstates of the Hamiltonian $\hat H\equiv
\frac{\hbar^2(q-q_{r})^2}{2m_*}-\frac{1}{2}m_*\omega_*^2 \nabla_q^2$, which 
may be calculated as discussed in Sec.~\ref{sec:RotonWavefunction}.

\subsection{Density modulations}
Note that density modulations are given by the $f_-$ amplitudes, in the form:
$\delta n ({\bf r,t})\equiv n({\bf r},t)-n_0({\bf r}) \propto \sqrt{n_0({\bf
r})} \Re(f_-({\bf r}))$.«, where $\Re$ denotes the real part. For the $\omega=0$
case, and since $W(\chi)\simeq 1$, from Eq.~\eqref{BdGa} one obtains
that $f_- \simeq (E(q)/\epsilon) f_+$.  For the $\omega>0$ case, since
$W(x)\simeq 1$ still, and we assume that $\tilde F({\bf q})$ is narrowly
peaked around $q_r$, we obtain the same result for the trapped case. Hence,
apart from a constant, $f_-({\bf r})\propto f_+({\bf r})\propto \psi_0({\bf
r})F(\boldsymbol{\rho})$.  As a result, $\delta n ({\bf r,t})\propto n_0({\bf
r}) \Re(F(\boldsymbol{\rho}))$.


\begin{thebibliography}{99}
\bibitem{footnote-T} In our calculations $a_i$ is chosen close to $a_c$, but its
actual value is not relevant, since we just employ the ground-state 
for $a=a_i$ and add numerical noise. However, the value of $a_i$ may be
relevant for the actual fluctuations. If the spectrum is weakly (or not)
rotonized for $a_i$, the initial population of the dominant unstable modes for
$a=a_f$ (which are around the roton minimum) is negligible when $k_B T\ll
\mu_l(0)$ ($k_B$ is the Boltzmann constant), corresponding to small $\xi$ in
Fig.~\ref{fig:2}. In contrast, the initial population of unstable modes may be
significant for $k_B T\sim \epsilon_r$, if the roton depth approaches
zero for $a_i$, corresponding to large $\xi$ in Fig.~\ref{fig:2}.
\bibitem{footnote-otherspecies} Similar results are expected for other magnetic
species such as chromium and dysprosium, although the particular parameters
will, of course, change.
\bibitem{footnote-k0} Prior to destabilization the dBEC presents a narrow
momentum distribution at $q=0$ with a width $\propto 1/R$.  We have removed the
$q=0$ contribution in  Figs.~\ref{fig:2}(b) and ~\ref{fig:2}(f), to highlight
the ring-like pattern as well as the azimuthal structure along the ring when
several $s$ states are mixed. A  thorough analysis of the momentum distribution
is performed in Sec.~\ref{sec:TOF}.
\end{thebibliography}
\end{document}